\documentclass[runningheads]{llncs}
\usepackage{graphicx}
\usepackage{cite}

\usepackage{hyperref}

\hypersetup{colorlinks,linkcolor={blue},citecolor={blue},urlcolor={blue}}

\usepackage{amsmath}
\usepackage{amsfonts}
\usepackage{subcaption}
\usepackage{amssymb}
\usepackage{mathtools}
\usepackage{enumitem}
\usepackage{soul}
\usepackage{color}

\usepackage{tcolorbox}
\newtcolorbox{mybox}{colback=gray!25!white,colframe=gray!25!white}

\newcommand{\iffullversion}[2]{%
#1%
}

\begin{document}

\title{A Simple LP-Based Approximation Algorithm for the Matching Augmentation Problem}
\titlerunning{A Simple LP-Based Approximation Algorithm for the MAP}

\author{\'Etienne Bamas \and
Marina Drygala \and
Ola Svensson}

\authorrunning{E. Bamas et al.}

\institute{EPFL, Switzerland}
\date{}

\maketitle
\begin{abstract}

    The Matching Augmentation Problem (MAP) has recently received significant attention as an important step towards better approximation algorithms for finding cheap $2$-edge connected subgraphs. This has culminated in a $\frac{5}{3}$-approximation algorithm.  However, the algorithm and its analysis are fairly involved and do not compare against the problem's well-known  LP relaxation called the cut LP.

    In this paper, we propose a simple algorithm that, guided by an optimal solution to the cut LP,  first selects a DFS tree and then finds a solution to MAP by computing an optimum augmentation of this tree.   
   Using properties of extreme point solutions, we show that our algorithm always returns (in polynomial time) a better than $2$-approximation when compared to the cut LP. We thereby also obtain an improved upper bound on the integrality gap of this natural relaxation.


   
   
   
   
\end{abstract}
\newcommand{\LP}{\textrm{LP}}
\newcommand{\Se}{\mathcal{S}}

\newcommand{\TODO}{\textcolor{red}{TODO}}


\section{Introduction}
Designing cheap networks that are robust to edge failures is a basic and important problem in the field of approximation algorithm. The area containing these problems is often referred to as \textit{survivable network design}. Generally, one has to compute the cheapest network that satisfies some connectivity requirements in-between some prespecified set of vertices. Classic examples are for instance the Minimum Spanning Tree problem in which one has to augment the connectivity of a graph from 0 to 1 or related questions such as the Steiner Tree/Forest problem. Another type of network design is to build 2-edge connected spanning subgraph (2-ECSS) or multisubgraph (2-ECSM), where one has to augment the connectivity of a graph from 0 to 2. The latter problems are closely related to the famous Traveling Salesman Problem (TSP). Unfortunately, most of the problems in this area are NP-hard (or even APX-hard), and what one can hope for is generally to compute an approximate solution in polynomial time. Powerful and versatile techniques such as \textit{primal-dual} \cite{goemans1995general,williamson2011design} or \textit{iterative rounding} \cite{jain2001factor,lau2011iterative} guarantee an approximation within factor 2 for many of these problems but improving on this bound for any connectivity problem is often quite challenging. In the case of 2-ECSS, a $4/3$-approximation is known if the underlying graph $G$ is unweighted \cite{DBLP:journals/combinatorica/SeboV14,4/3ECSS}. However, a similar result for the weighted case has remained elusive, and the best approximation algorithm only guarantee a factor 2 approximation. A prominent special case of the weighted 2-ECSS problem is the so-called Forest Augmentation Problem (FAP). In such instances of 2-ECSS all edge weights are either 0 or 1 (we will refer to edges of cost 0 as \textit{light edges} and edges of cost 1 as \textit{heavy edges}). The name stems from the fact that one can assume that the light edges form a forest $F$, and the goal is to find the smallest set of heavy edges $E'$ such that $F\cup E'$ is 2-edge connected.

A famous special case of FAP is the Tree Augmentation Problem (TAP) which has been extensively studied for decades. In this problem, the forest $F$ is a single spanning tree, and one has to find the smallest set of edges to make the tree 2-edge connected. For this problem, several better-than-2 approximations were designed in a long line of research \cite{khuller1993approximation,frederickson1981approximation,nagamochi2003approximation,kortsarz2018lp,kortsarz2015simplified,cohen20131+,cheriyan2018approximating2,even20091,cheriyan2018approximating,cheriyan2008integrality,cecchetto2021bridging,adjiashvili2018beating,fiorini2018approximating,traub1,traub2,nutov2021tree,grandoni2018improved}. One can see TAP as an extreme case of FAP where the forest is a single component. Another interesting special case is the Matching Augmentation Problem (MAP), in which the forest of light edges forms a matching $M$ and one has to find the smallest set of heavy edges $E'$ such that $M\cup E'$ is 2-edge connected. It can be seen as the other extreme case in which the forest forms as many components as possible. We also remark that MAP generalizes the unweighted 2-ECSS problem, which can be viewed as an instance of MAP with an empty matching. For MAP, only recently a better-than-2 approximation was given by Cheriyan et al. \cite{cheriyan2020matching,cheriyan2020improved}. These two works culminate in a $5/3$-approximation, obtained via a fairly involved algorithm and analysis. 

For many of these network design problems, there is a simple linear programming relaxation called the cut LP. In the case of FAP, for a given graph $G=(V,E)$, forest $F\subseteq E$ the cut LP is written as follows, with a variable $x_e$ to decide to take each edge $e$ or not. Recall that $\delta(S)$ denotes the edges with exactly one endpoint in $S$.

\begin{align*} \label{LP:cut_LP}
    LP(G,F):  & \quad \min \sum_{e\in E\setminus F} x_e   \\
    & \quad\sum_{e\in \delta(S)} x_e \ge 2,\quad  \text{for all }S, \emptyset \subsetneq S\subsetneq V \\
    & \quad 0\leq x_e \leq 1, \quad \quad \quad \forall e\in E.
\end{align*}

The integrality gap of this linear program is an interesting question by itself. Recently, in the case of TAP (i.e. $F$ is a spanning tree), Nutov \cite{nutov2021tree} showed that the integrality gap is at most $2-2/15\approx 1.87$. Cheriyan et al. \cite{cheriyan2008integrality} showed that the integrality gap is at least $3/2$ in the case of TAP. In the case of MAP, the best upper bound on the integrality gap is 2, and the best lower bound is $9/8$ \cite{DBLP:journals/combinatorica/SeboV14,alexander2006integrality}. We note that the recent works \cite{cheriyan2020matching,cheriyan2020improved} do not seem to compare against the cut LP, and therefore do not show an integrality gap better than 2 for MAP.

\subsection{Our results}

In this paper, we give an algorithm that guarantees an approximation ratio $2-c$ (for some absolute constant $c>0$) with respect to the best fractional solution of the cut LP. The algorithm is the following. We note that some of our techniques are reminiscent of the algorithm of Mömke and Svensson \cite{momke2016removing} for the travelling salesman problem (see also \cite{DBLP:journals/siamdm/Newman20,DBLP:journals/mst/Mucha14} for follow-up works).

\vspace{1mm}
\begin{mybox}
\textbf{The LP-based algorithm:} 
\begin{enumerate}
    \item Compute an optimal extreme point solution $x^*$ to $LP(G,M)$.
    \item Let $E'=\{e\in E, x_e^*>0\}$ be the \textit{support} of $x^*$, and run a DFS on the support graph $G'=(V,E')$ which always give priority first to an available light edge and second to the available heavy edge $e$ maximizing $x^*_e$.
    \item Compute an optimum augmentation $A$ to the TAP problem with respect to the DFS tree $T$ computed in the previous step and return $H=T\cup A$.
\end{enumerate}
\end{mybox}

We note that the LP-based algorithm indeed runs in polynomial time. Step 2 computes a DFS in which some edges are explored in priority (if possible). Step 3 can also be done in polynomial time because the tree $T$ is a DFS tree. This implies that all non-tree edges are back-edges (i.e. one endpoint is an ancestor of the other). In the language of TAP, these edges are often referred to as ``uplinks", and it is well-known that TAP instances in which the edges are only ``uplinks" are solvable in polynomial time \cite{cohen20131+,frederickson1982relationship}. 

Finally, the solution given by the algorithm is feasible since Step 2 increases connectivity from 0 to 1 and Step 3 from 1 to 2. One can check that no edge is taken twice in the process since $A$ and $T$ are disjoint.

In this paper, our main result shows that this simple algorithm guarantees an approximation within factor strictly better than 2 with respect to the cut LP relaxation.

\begin{theorem}\label{thm:main}
The LP-based algorithm returns a feasible solution to any MAP instance of cost at most $2 - c$ times the cost of the fractional solution $x^*$, for some absolute constant $c>0$.
\end{theorem}
For the sake of exposition, we did not try to optimize the constant $c$ but we believe that improving the ratio of $5/3$ in \cite{cheriyan2020improved} (that holds with respect to the optimum \textit{integral} solution) would require new techniques in the analysis. Since Nutov \cite{nutov2021tree} proved the integrality gap of the cut LP to be strictly better than 2 for TAP, the cut LP seems a promising relaxation for the general FAP. Additionally, we prove the following simple theorem.

\begin{figure}[ht]
\begin{subfigure}{.5\textwidth}
  \centering
   \includegraphics[scale=0.95]{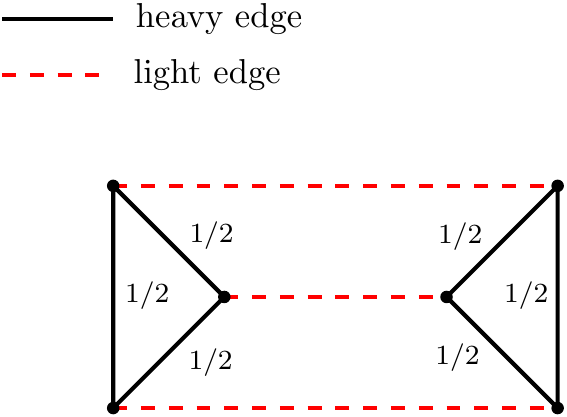}
    \caption{An integrality gap example}
    \label{fig:gap}
\end{subfigure}
\begin{subfigure}{.5\textwidth}
  \centering
  \includegraphics[scale=0.7]{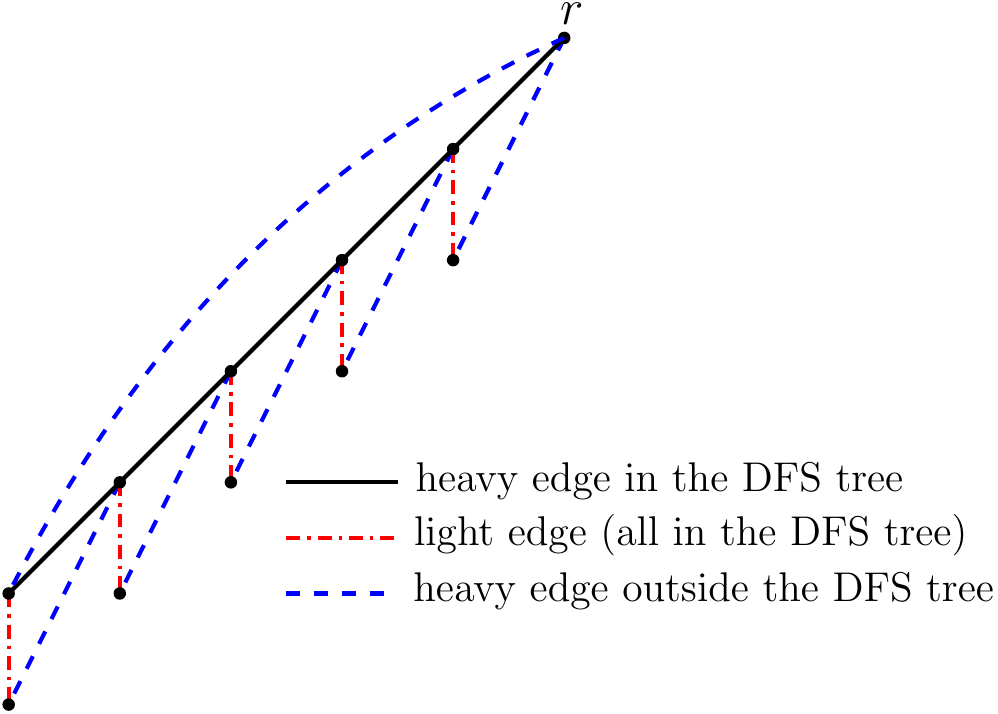}
    \caption{An example of a bad DFS tree.}
    \label{fig:badDFS}
\end{subfigure}
\end{figure}

\begin{theorem}
The integrality gap of the cut LP for MAP is at least $4/3$.
\end{theorem}
\begin{proof}
Consider the example given in Figure \ref{fig:gap}, which is a simple adaptation of a classic example for the related TSP problem. One can check that the fractional solution that gives $1/2$ fractional value to all heavy edges and value $1$ to all light edges is feasible for a total cost of $6/2=3$. However, any integral solution costs at least 4.
\end{proof}

\subsection{Our techniques}

The proof of Theorem \ref{thm:main} relies on several crucial observations that we sketch here. The first observation is that the total cost of the DFS tree $T$ is always at most the cost of $x^*$ (denoted $c(x^*)$). This follows because $T$ must contain all the light edges since they are given priority over any other edge (note that since we assume that $M$ is a matching, it cannot happen that two distinct light edges want priority at the same time). Therefore, the total cost of the tree $T$ is exactly equal to $n-1-|M|$, while it is easy to show that $c(x^*)\geq (n-|M|)$.

 Another interesting fact is that if one considers the LP solution $x^*$ restricted to the edges \textit{not} in the tree $T$ (denote this solution by $x^*_{E\setminus T}$), then this is a feasible solution to the cut LP of the TAP instance with respect to the tree $T$ (i.e. $x^*_{E\setminus T}$ is a feasible solution to $LP(G,T)$). Hence, if we denote by $y^*$ the optimum fractional solution to $LP(G,T)$, we have that $c(y^*)\leq c(x^*_{E\setminus T})$.

 Because $T$ is a DFS tree, the TAP instance with respect to the tree $T$ contains only ``uplinks'' and therefore $LP(G,T)$ is known to be integral \cite{adjiashvili2018beating}. We note that this already gives a simple proof that the integrality gap of $LP(G,M)$ is at most 2. To get better than 2, we only need to show that 
\begin{equation*}
    (n-|M|-1)+c(y^*)\leq (2-c)c(x^*).
\end{equation*}
Conceptually, we distinguish between two cases. If $c(x^*)>(1+c)(n-|M|)$ (i.e. the LP solution is expensive), then the DFS tree is significantly cheaper than $c(x^*)$ and it is easy to conclude that the cost of our solution $T\cup A$ is better than $2c(x^*)$. 
Otherwise, assume that the LP value is close to the trivial lower bound of $(n-|M|)$. In this case, we show that $c(y^*)\leq (1-c)c(x^*)$.

To show this, we consider two possibilities. We can prove that either we can scale down a significant portion of $x^*_{E\setminus T}$ to obtain a cheaper feasible solution to $LP(G,T)$, or that $c(x^*_{E\setminus T})$ itself is already significantly smaller than $c(x^*)$. When a lot of the tree cuts in $T$ (i.e. the cuts defined by removing an edge from $T$ to obtain two trees and taking the edges with one endpoint in each tree) have some slack in the TAP solution $x^*_{E\setminus T}$ (that is when a lot of tree cuts $S$ satisfy $x^*_{E\setminus T}(\delta (S))>1+c$), the first case is realized. Otherwise, when almost all of the tree cuts are nearly tight (i.e. satisfy $x^*_{E\setminus T}(\delta (S))\leq 1+c$), we can show that the DFS must have captured a good fraction of the value of $c(x^*)$ \textit{inside} the tree $T$. This step uses some crucial properties of extreme point solutions as well as our choice of DFS. Therefore the cost of $x^*_{E\setminus T}$ is significantly smaller than the cost of $x^*$ completing the argument.

Before proceeding to the proof, it is worthwhile to mention that we are not aware of any example on which our algorithm has a ratio worse than $4/3$ times the cost of $x^*$. It remains open to give a tighter analysis of this algorithm. We also note that \cite{KV94} also makes use of DFS for the related problem of unweighted 2-ECSS. They obtain a ratio of $3/2$ for the unweighted 2-ECSS problem. However, their DFS is not LP-based and we remark that if we do not guide the DFS with the LP solution, the approximation ratio can be arbitrarily close to 2. We give an example in Figure \ref{fig:badDFS}. One can see that the DFS tree (rooted at $r$) contains all the matching edges, and the tree augmentation problem requires us to take all but one of the back-edges. However, the optimum solution to the MAP instance is to take a Hamiltonian tour containing all the light edges. Generalizing the same example by simply increasing the depth of the tree leads to an approximation arbitrarily close to 2.


\section{The Analysis of the LP-based Algorithm}
In this section, we prove Theorem \ref{thm:main}. It is organized as follows. In subsection \ref{subsec:prelimproof}, we introduce some basic definitions. In the subsequent subsection, we proceed via a case distinction to prove the theorem.

\subsection{Preliminaries}
\label{subsec:prelimproof}
We will use $T$ to refer to the DFS tree computed by the algorithm, and we will list edges in $G$ as $uv$, where $u$ is an ancestor of $v$ in $T$. Since $T$ is a DFS tree, all edges in $G$ must have the property that one endpoint is an ancestor of the other in $T$. We will let $B=E \setminus T$ denote the set of \textit{back-edges} of $G$.
As in the introduction, we will call an edge of weight 1 a \textit{heavy edge} and an edge of weight 0 a \textit{light edge}. For every edge $e$ in the DFS tree $T$ computed, we let $T(e)$ denote the tree cut corresponding to the edge $e$ in the tree $T$. Formally, $T(e)= \delta(T_v)$, where $e=uv$ and $T_v$ is the sub-tree rooted at $v$. We call an edge $e\in T$ $\alpha$-\textit{tight} if we have 
\begin{equation*}
    x^*(T(e))-x_e^*< 1+\alpha. 
\end{equation*}
Implicitly, if we call an edge $e$ $\alpha$-tight, this will mean that $e$ belongs to the tree $T$. In addition, we denote by $N_t^{(\alpha)}$ the number of $\alpha$-tight edges in the tree $T$. For a tree $T$, we denote by $x^*_T$ the restriction of $x^*$ to the edges in the tree $T$. We note that for any instance of the MAP, it must be that $c(x^*)\geq (n-|M|)$. This follows by a simple double counting argument on the fractional degree of each component (precisely we have $n-|M|$ components that must have fractional degree 2 each). It is also clear that the DFS tree $T$ must contain all the light edges in $M$ since they are given priority. Hence the cost of $T$ is at most $n-|M|-1\leq c(x^*)$. In the following, we will fix two parameters $\epsilon=10^{-1},\gamma=10^{-3}$.

\subsection{The analysis of the algorithm}
We note that if $c(x^*)\geq (1+\gamma) (n-|M|)$, it is easy to show that the cost of the returned solution $T\cup A$ is at most 
\begin{equation}
\label{eq:bound1}
    (n-|M|-1)+c(x^*)\leq \frac{c(x^*)}{1+\gamma}+c(x^*)=c(x^*)\left( 2-\frac{\gamma}{1+\gamma}\right).
\end{equation}
However, if $c(x^*)< (1+\gamma) (n-|M|)$ and $N_t^{(\gamma)}\leq (1-\gamma)(n-|M|)$ (i.e. there are few $\gamma$-tight tree cuts), then we proceed as follows.
We partition the set of back edges in our graph $B$ into $B_t^{(\gamma)} \cup B_s^{(\gamma)}$, where $B_t^{(\gamma)}$ contains all edges $e \in B$ that are contained in $T(e')$ for some $\gamma$-tight edge $e' \in T$. Then $x'$, defined by $$x'(e) = \begin{cases} x^{*}(e) & e \in B_t^{(\gamma)} \\ \frac{x^{*}(e)}{1+\gamma} & e \in B_s^{(\gamma)}\\ 1 & \text{ otherwise} \end{cases}$$ is also a feasible solution to $LP(G,T)$. The total fractional value represented by edges in $B_t^{(\gamma)}$ is at most $(1+\gamma) N_t^{(\gamma)}$. Hence, $c(x')$ can be upper bounded as follows.
\begin{equation*}
    c(x')\leq \frac{c(x^*)-(1+\gamma) N_t^{(\gamma)}}{1+\gamma} + (1+\gamma) N_t^{(\gamma)} = \frac{c(x^*)}{1+\gamma}+\gamma N_t^{(\gamma)}.
\end{equation*}
Since the cost of $T\cup A$ is at most $c(x^*)+c(x')$ and we assume that $N_t^{(\gamma)}\leq (1-\gamma)(n-|M|)$, it is easy to get the upper-bound of 
\begin{equation}
\label{eq:bound2}
    c(x^*)\left(1 + \frac{1}{1+\gamma}\right)+\gamma (1-\gamma)(n-|M|)\leq c(x^*)\left(2-\frac{\gamma^3}{1+\gamma} \right),
\end{equation}
where the last inequality follows because $n-|M| \leq c(x^*)$.
Since these two cases clearly give a better than 2 approximation, we assume in the rest of the analysis that 
\begin{equation}
\label{eq:assumption1}
    (n-|M|)\leq c(x^*)< (1+\gamma) (n-|M|),
\end{equation}
and 
\begin{equation}
    \label{eq:assumption2}
    N_t^{(\gamma)}> (1-\gamma)(n-|M|).
\end{equation}
We will show that $c(x^{*}_T)$ is at least a constant fraction times $c(x^{*})$. Since the cost of the returned solution $T\cup A$ is at most $2c(x^*)-c(x_T^*)$, this will conclude the proof. First, we partition the $\gamma$-tight tree cuts into two sets of cuts $\Se_0$ and $\Se_1$ containing the tight tree cuts associated with light edges and heavy edges, respectively. We can then distinguish between two sub-cases. For each edge $e=uv\in T$, we say that $e$ is a \textit{leaf} edge if $v$ is a leaf in the tree $T$ (recall that we always write an edge $e$ as $e=uv$ such that $v$ is a descendant of $u$ in $T$). We denote $\Se_0^+$ the non-leaf edges in $\Se_0$ and $\Se_0^-$ the leaf edges in $\Se_0$. We have two main cases.

\subsubsection{Suppose that $|\Se_1|\geq \gamma (n-|M|)$ or that $|\Se_0^+|\geq \gamma (n-|M|)$.} 
By feasibility of $x^{*}$ at least 2 units of $x^*$ must cross any tree cut. Hence $x^{*}(\delta(T_v)) \geq 2$, for any $v\in V$. By definition of $\gamma$-tightness we know that for any $\gamma$-tight edge $e=uv$ we have $x^{*}(e) \geq x^{*}(\delta(T_v)) - (1+\gamma)\geq 1-\gamma$. 

Hence if $|\Se_1|\geq \gamma (n-|M|)$, we have that 
\begin{equation*}
    c(x_T^*)\geq \gamma(1-\gamma) (n-|M|)\geq \frac{\gamma(1-\gamma)}{1+\gamma}c(x^*),
\end{equation*}
which concludes the case when $|\Se_1|$ is large. In the following we use some properties of extreme point solutions. We say that an edge $e$ is \textit{fractional} (with respect to the fractional solution $x^*$) if $0<x_e^*<1$. A vertex $v$ is said to be $\alpha$-\textit{fractional} if it has more than $1/\alpha$ incident fractional edges in the support of $x^*$ (for any $\alpha>0$). We claim the following lemma, the proof of which relies on standard techniques and can be found in \iffullversion{Appendix \ref{sec:appendix_proofs}}{the full version of this paper}. We note that a similar result was used in \cite{momke2016removing}.
\begin{lemma}
\label{lem:support}
If $x^*$ is an extreme point solution of the cut LP, then there are at most $2n-1$ fractional edges in $G$. Moreover, for any $\alpha>0$, there are at most $4\alpha n$ $\alpha$-fractional vertices with respect to $x^*$.
\end{lemma}

Using Lemma \ref{lem:support} with $\alpha=\gamma/16$, we get that if $|\Se_0^+|\geq \gamma (n-|M|)$, then (recall that $n-|M|\geq n/2$) there are at least 
\begin{equation*}
    \gamma (n-|M|) - (\gamma/4)n\geq (\gamma/2)(n-|M|)
\end{equation*}
edges $uv\in \Se_0^+$ such that $v$ is not $\gamma/16$-fractional. We then claim the following simple lemma.

\begin{lemma}\label{lem:node_cut}
Fix any $\alpha,\alpha'>0$. Suppose that $e=uv$ is an $\alpha$-tight light edge, such that $v$ is not a leaf in $T$. Then if $v$ is not $\alpha'$-fractional there exists some edge $e'=vw$ in $T$ such that $x^{*}(e') \geq (1-\alpha) \alpha'$.
\end{lemma}
\begin{proof}
By feasibility of $x^{*}$ we know that $x^{*}(\delta(T_v \setminus v)) \geq 2.$ Since $e$ is $\alpha$-tight and $T$ is a DFS tree we have that $x^{*}(\delta(T_v)) - x^{*}(e) \leq 1+\alpha$. We know that $E(T_v \setminus v, v) = \delta(T_v \setminus v) \setminus \delta(T_v)$, and as a result $x^{*}(E(T_v\setminus v,v)) \geq 1 - \alpha$. Since $v$ is not $\alpha'$-fractional there must be an edge $e' \in E(T_v\setminus v,v)$ with value at least $x^{*}(E(T_v\setminus v,v)) \alpha'\geq (1-\alpha)\alpha'$. Since our DFS selects always the highest possible fractional value if there is no light edge to explore, the first edge selected after exploring $v$ must be of fractional value at least $(1-\alpha)\alpha'$.
\qed
\end{proof}

Combining Lemma \ref{lem:node_cut} with the previous observation, if $|\Se_0^+|\geq \gamma (n-|M|)$ we get that
\begin{equation*}
    c(x_T^*)\geq (\gamma/2)(n-|M|)(1-\gamma) (\gamma/16)\geq c(x^*)\frac{\gamma^2(1-\gamma)}{32(1+\gamma)}.
\end{equation*}
Combining these two cases we get that if $|\Se_1|\geq \gamma (n-|M|)$ or $|\Se_0^+|\geq \gamma (n-|M|)$ then 
\begin{equation*}
    c(x_T^*)\geq c(x^*)\cdot \min\left( \frac{\gamma^2(1-\gamma)}{32(1+\gamma)},\frac{\gamma(1-\gamma)}{1+\gamma}\right),
\end{equation*}
hence the cost of $T\cup A$ is upper bounded by 
\begin{equation}
\label{eq:bound3}
    2c(x^*)-c(x_T^*)\leq c(x^*)\left(2-\frac{\gamma^2(1-\gamma)}{32(1+\gamma)}\right),
\end{equation}
which is clearly better than 2. Hence we are left with the last case, in which 
\begin{equation*}
    |\Se_0^-|> (1-\gamma)(n-|M|)-|\Se_1|-|\Se_0^+|> (1-3\gamma)(n-|M|).
\end{equation*}

\subsubsection{Suppose $|\Se_0^-|> (1-3\gamma)(n-|M|)$.}

This is the most interesting case. Note that for each edge $e=uv\in \Se_0^-$, the fractional degree of $v$ restricted to heavy edges must be at least 1, and all of this fractional degree is carried by backedges in $T$. Denote by $B'$ this subset of backedges. Next we define $B''\subseteq B'$ to be the subset of $B'$ containing only edges with fractional value at least $\epsilon=10^{-1}$. We claim that 
\begin{equation}
\label{eq:Bbound}
    |B''|\geq n/10. 
\end{equation}

Assume the contrary, since the fractional value of any edge is at most 1 then the total value carried by edges in $B'\setminus B''$ must be at least
\begin{equation*}
    |\Se_0^-|-(n/10)>(1-3\gamma)(n-|M|)-n/10>3n/8-n/10.
\end{equation*}
(Recall that $(n-|M|)\geq n/2$ and $(1-3\gamma)>3/4$). Since all the edges in $B'\setminus B''$ have fractional value at most $\epsilon$, there must be at least $(3n/8-n/10)/\epsilon = 30n/8-n > 2n-1$ such edges, contradicting Lemma \ref{lem:support}. Hence $|B''|\geq n/10$.

For completeness we consider the case when $E$ contains heavy edges that are parallel to light edges. Partition $B''$ into $B''_1 \cup B''_2$, where $B''_2$ is the set of edges in $B''$ parallel to an edge in $\Se_0^-$. We define $B''_1$ to be the remaining edges in $B''$.

We claim that $|B''_2| \leq n/100$, and thus loosely $|B''_1| \geq n/20$.

To see this note that,
\begin{equation}
\label{eq:surplusparallel}
    c(x^*)-(n-|M|)\geq |B''_2|\epsilon.
\end{equation}

Equation \eqref{eq:surplusparallel} holds as the lower bound of $(n-|M|)$ on $c(x^*)$ is obtained only by counting the fractional degree of each component in $M$. Since those parallel edges are not counted in this bound (they are only \textit{within} a single component), they directly count in the value of $c(x^*)-(n-|M|)$, which counts the surplus of $c(x^*)$ above $(n-|M|)$.


Then as $c(x^*)-(n-|M|) \leq \gamma (n-|M|)$, by choice of $\epsilon$ and $\gamma$ we obtain that $|B''_2| \leq n/100$.

 Consider the set of vertices $X$ that contains the ancestor vertices of the edges in $B''_1$. We claim that 
\begin{equation}
    |X|\geq n/500.
\end{equation}
To prove this, we first claim that 
\begin{equation}
\label{eq:surplusmatching}
    c(x^*)-(n-|M|)\geq |B_1''|\epsilon - 2|X|.
\end{equation}
To see this, note again that the value $c(x^*)-(n-|M|)$ represents the surplus value of $c(x^*)$ above the lower bound that gives fractional degree 2 to every vertex. This trivial lower bound gives a fractional value---which is the fractional degree restricted to heavy edges---of at most 2 to every vertex, hence a fractional value of at most $2|X|$ to the set of vertices $X$. Since every edge in $B_1''$ has fractional value of at least $\epsilon$ and is adjacent to a single vertex in $X$, we get that the surplus value of $c(x^*)$ above the trivial lower bound is at least $|B_1''|\epsilon - 2|X|$ which proves Equation \eqref{eq:surplusmatching}.

Since by assumption we have $c(x^*)<(1+\gamma)(n-|M|)$ we conclude with Equation \eqref{eq:surplusmatching} that
\begin{equation*}
    \gamma (n-|M|)>c(x^*)-(n-|M|)\geq |B_1''|\epsilon - 2|X|
\end{equation*} which implies, by our lower bound on $|B_1''|$ and our choice of $\gamma$ and $\epsilon$,
\begin{equation}
    |X|\geq \frac{|B_1''|\epsilon-\gamma (n-|M|)}{2} \geq \frac{n/200-n/10^3}{2}=n/500.
\end{equation}

For each vertex $u\in X$, denote by $e_u$ the first edge selected by the DFS after reaching $u$. Denote $X'\subseteq X$ the subset of $X$ containing only vertices $u\in X$ such that $e_u$ does \textit{not} belong to $\Se_0^+$. Then, we have by assumption,
\begin{equation*}
    |X'|\geq |X|-|\Se_0^+| \geq n/500 - \gamma (n-|M|) \geq n/500-n/10^3=n/10^3.
\end{equation*}
We finally claim the following, which crucially uses how the DFS selects the edges to explore in priority.
\begin{claim}
\begin{equation*}
    c(x^*_T)\geq \epsilon |X'|.
\end{equation*}
\end{claim}
\begin{proof}
    There are two cases to consider (depicted in Figure \ref{fig:leaves}). 
    
    \begin{figure}
        \centering
        \includegraphics[scale=0.8]{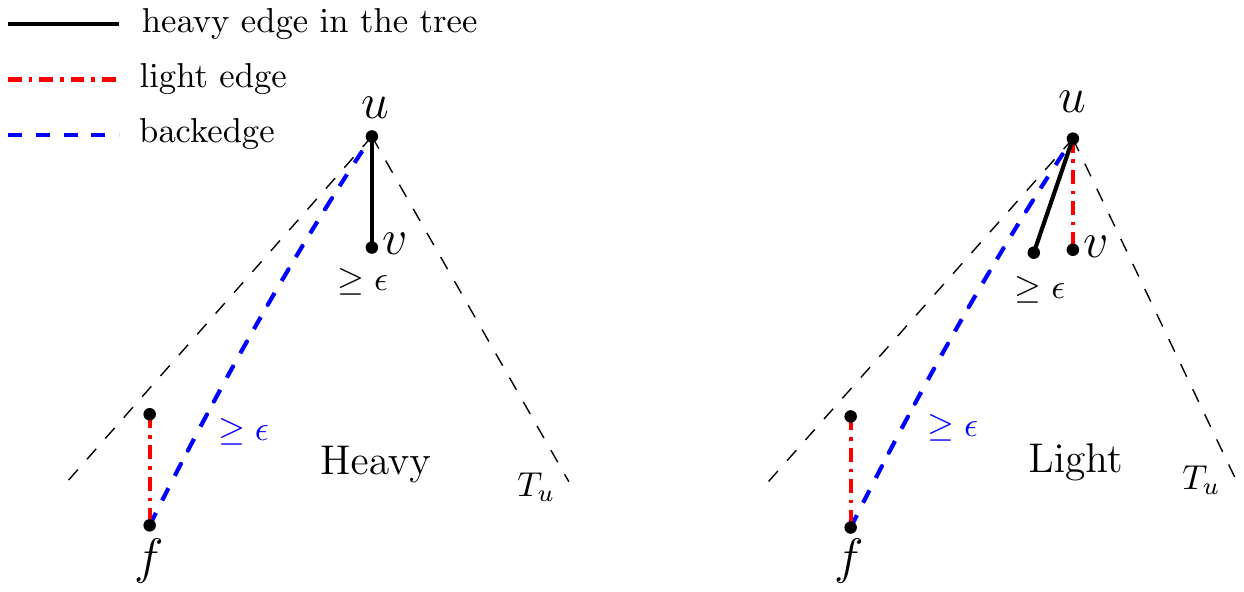}
        \caption{On the left side, the case when the first edge selected out of $u$ is heavy. On the right the case when the first edge selected out of $u$ is light.}
        \label{fig:leaves}
    \end{figure}
    
    If $u\in X'$ is such that $e_u=uv$ is a heavy edge, by definition of $X'$ there must be an edge $e'=uf$ coming from a leaf $f$ in the tree $T$ to $u$ of fractional value $x_{e'}\geq \epsilon$. At the first time the DFS visits the vertex $u$, the leaf $f$ was not explored yet hence the edge $e'$ was a valid choice of edge to explore. Since our DFS always takes the highest fractional value, it must be that 
    \begin{equation*}
        x_{e_u}\geq x_{e'}\geq \epsilon.
    \end{equation*}

If $u\in X'$ is such that $e_u=uv$ is a light edge, recall that by definition of $X'$, $v$ must be a leaf in $T$. Then when the DFS arrived at $v$, it must be that all reachable vertices from $v$ were already visited. Hence the DFS must have backtracked to $u$. Now note that by our construction of $B_1''$, we know that there must be another leaf $f$ such that $e'=uf$ is a back-edge in the tree of fractional value $x_{uf}\geq \epsilon$ (recall that $uf$ is not parallel to the edge $e_u$). Since $f$ is a leaf, $u$ must have been explored before $f$ therefore, after backtracking from $v$ to $u$ the edge $uf$ was a valid edge to take. Therefore the DFS must have selected a second edge $e''$ in the tree from $u$ such that 
\begin{equation*}
    x_{e'}\geq x_{uf}\geq \epsilon.
\end{equation*}

Hence we proved that all vertices $u$ in $X$ must be adjacent to at least one heavy edge of fractional value $\epsilon$ that belongs to the tree $T$ and goes to a child of $u$. Hence the proof of the claim.
\qed
\end{proof}

By the previous claim, we have $c(x^*_T)\geq \epsilon |X'|$ hence the cost of the returned solution $T\cup A$ is at most 
\begin{equation}
    2c(x^*)-c(x^*_T)\leq 2c(x^*) - n/10^4 \leq c(x^*)\left(2-10^{-4} \right),
\end{equation}
which ends the proof of Theorem \ref{thm:main}.

\section{Conclusion}
In this paper, we gave a simple $2-c$ approximation algorithm for MAP with respect to the standard cut LP. Our algorithm computes a DFS tree using an optimal extreme point solution to the above-mentioned LP solution as a guide when selecting edges and then augments the resulting tree optimally. We leave it as an open problem to see if the analysis can be refined to get an improved guarantee for the algorithm. We remark that it is not difficult to see that if the LP solution is $f$-fractional, then our algorithm produces a solution of at most $2-f$ times its value. In particular, this gives an upper bound of $3/2$ for half-integral solutions. We wonder if a better understanding of the algorithm will lead to a $\frac{4}{3}$-approximation for the half-integral case.

Another interesting connection of our work to related works is by its relevance to the Path Augmentation Problem (PAP). An instance of PAP is an instance of FAP where the forest $F$ contains only paths. 
We note that our techniques generalize to instances of PAP, if the cut LP returns a solution of cost equal to the number of components in $F$. This follows because the support of the optimal extreme point solution of the cut LP for such instances has no fractional value incident to internal nodes of the paths in $F$.
Some independent work \cite{grandoni2021breaching} shows that the general FAP reduces to special instances of PAP where the cost of the LP solution is almost equal to the number of components in $F$. However these techniques do not preserve the integrality gap. Determining whether we can bound the integrality gap of the cut LP for the FAP strictly below 2 remains an interesting open problem.


\iffullversion{
\appendix
\section{Deferred proofs}
\label{sec:appendix_proofs}
Suppose that $x^{*}$ is an extreme point solution of $LP(G, M)$. We know that $x^*$ can be defined as the unique solution to the following system of $|E|$ equations, for some $\mathcal{S} \subseteq 2^V$ and $E_0 \cup E_1 \subseteq E$.

\begin{align*} \label{tight_extreme_point_constraints}
    & \sum_{e\in \delta(S)} x_e = 2,\quad  \text{for all }S \in \mathcal{S} \\
    &  x_e = 0 \quad \quad \quad \forall e\in E_0\\
    &  x_e = 1 \quad \quad \quad \forall e\in E_1\\
\end{align*}

 Lemma \ref{lem:laminar_family} shows that we can select $\mathcal{S}$ not too large. The proof of this lemma is the same as Theorem 4.9 from \cite{cornuejols1985traveling}.

\begin{lemma}[Theorem 4.9 in \cite{cornuejols1985traveling}]\label{lem:laminar_family} Let $x^{*}$ be an extreme point of the MAP cut LP then the family of equations $\mathcal{S}$ can be chosen to be a laminar family. 
\end{lemma} 

It is well known that any laminar family has size at most $2n-1$. Therefore the number of fractional edges is at most $|E|-|E_0|-|E_1|=|\Se|\leq 2n-1$.

}{}

\bibliographystyle{splncs04}
\bibliography{references}
\end{document}